# Reframing Convergent and Divergent Thought for the 21st Century

Liane Gabora (liane.gabora@ubc.ca)
Department of Psychology, University of British Columbia
Kelowna BC, V1V 1V7, CANADA

**Abstract**

Convergent and divergent thought are promoted as key constructs of creativity. *Convergent thought* is defined and measured in terms of the ability to perform on tasks where there is one correct solution, and *divergent thought* is defined and measured in terms of the ability to generate multiple solutions. However, these characterizations of convergent and divergent thought presents inconsistencies, and do not capture the reiterative processing, or 'honing' of an idea that characterizes creative cognition. Research on formal models of concepts and their interactions suggests that different creative outputs may be projections of the same underlying idea at different phases of a honing process. This leads us to redefine convergent thought as thought in which the relevant concepts are considered from *conventional contexts,* and divergent thought as thought in which they are considered from *unconventional contexts.* Implications for the assessment of creativity are discussed.

## Introduction

Other species perceive, make decisions, and take action, but our ability to adapt ideas to our own needs, tastes, and perspectives, and express ourselves through language, art, technology, and other means, is exceptional. Thus, understanding creative thinking is central to understanding our humanness.

In creativity research, as in other areas of cognitive science, there is a long history of dual process theories, which assert that there are two kinds of thought, or that thought varies along a continuum between two extremes (Evans & Frankish, 2009; James, 1890/1950, Sloman, 1996). In the creativity literature the distinction is usually made between convergent and divergent thinking[1]. *Convergent thought* is defined and measured in terms of the ability to perform on tasks where there is a single correct solution, while *divergent thought* is defined and measured in terms of the ability to generate multiple different solutions (Guilford, 1967). A widely used test of convergent thinking is the Remote Associates Test (RAT) (Mednick, 1968). A typical RAT question is: What is the common associate of TANK, TABLE, and HILL? The answer is: TOP. A widely used divergent thinking test is the Alternate Uses task, which asks questions like 'think of as many uses as you can for a brick' (Christensen, Guilford, Merrifield, & Wilson, 1960). Responses are most often rated in terms of *fluency,* the total number of ideas generated in a given time. Often they are additionally rated in terms of *originality,* the number of unusual or statistically infrequent ideas. Fluency and originality are considered to reflect the quantity and quality of ideation performance, respectively. Occasionally they are also rated in terms of *flexibility,* the number of different categories of ideas. On rare occasions answers are rated for *elaboration:* the amount of detail given, or evidence that the individual has followed an associative pathway for some distance.

Although these characterizations of convergent and divergent thought have stuck for half a century, as formulated, they present inconsistencies. For example, it is often said that a creatively demanding problem requires both convergent and divergent thought (e.g., Beersma & De Dreu, 2005; Gibson, Folley, & Park, 2009; Kerr & Murthy, 2004). However, given that convergent and divergent thought are defined in terms of the number of correct solutions, this makes no sense. A problem either has one correct solution or it has many; it cannot have both one and many. Moreover, the way convergent and divergent thought have been defined and measured is inconsistent with how people think about creativity; for example, although divergent thinking is thought to be the most promising candidate for the foundation of creative ability (Plucker & Renzulli, 1999; Runco, 2007), performance on the RAT would seem to be a better indicator of creativity than many tasks that would be classified as a divergent thinking task, such as 'list as many things as you can that are red'. Finally, it is often noted that earlier responses on a divergent thinking task are less creative than latter ones (Beaty & Silvia, 2012), but if divergent thinking is characterized in terms of the number of possible responses, this is the opposite of what one should expect, because with each response one gives, the number of remaining possible responses decreases by one. Thus, the conventional view would predict that, as one proceeds, one should start thinking more *convergently,* not more divergently.

More fundamentally, as noted elsewhere (Piffer, 2012), divergent thinking research, and creativity research in general, emphasizes the generation of multiple ideas over what is sometimes called *honing*—recursively reflecting on a question or idea by viewing it from different perspectives with the output of each such reflection providing the input to the next (Gabora, 2007, 2017). One thereby comes to a deeper, more nuanced understanding of it. Honing differs from elaboration in that it does not include additions or modifications to the idea that are tacked on willy-nilly; it refers specifically to modifications that arise in response to an overarching conceptual framework that is shepherding[2] the creative process. The structure of this overarching framework reflects the individual's *worldview:* their self-organizing web of understandings about their world and their place in that world (in other words, the creator's mind as experienced 'from the inside').

---
[1] Sometimes the distinction is between associative and analytic thought (e.g., Chrusch, C. & Gabora, L., 2013), or executive versus generative (e.g., Ellamil, Dobson, Beeman, & Christoff, 2012). See (Sowden et al., 2014) for how convergent and divergent thinking relate to other dual process theories.

Like other self-organizing systems, a worldview continually interacts with and adapts to its environment to minimize internal *entropy,* a measure of uncertainty and internal disorder. Hirsh, Mar, and Peterson (2012) use the term *psychological entropy* to refer to anxiety-provoking uncertainty, which they claim humans attempt to keep at a manageable level. Noting that uncertainty can be experienced not just negatively as anxiety but also positively as a wellspring for creativity (or both), the term psychological entropy has been expanded to refer to *arousal*-provoking uncertainty. Redefining psychological entropy in terms of arousal rather than anxiety is consistent with findings that creative individuals exhibit greater openness to experience and higher tolerance of ambiguity (Feist, 1998), which could predispose them to states of uncertainty or worldview inconsistency (Gabora, 1999). Their higher variability in arousal (Martindale & Armstrong, 1974) reflects a predisposition to invite situations that increase psychological entropy, experience them positively, and resolve them. In this way, psychological entropy—a macro-level variable acting at the level of the worldview as a whole—generates emotions that play a role in guiding and monitoring creative tasks.

Thus, honing continues until psychological entropy decreases to an acceptable level. In Piagetian terms, during honing the individual assimilates each new understanding of the idea, and the individual's worldview changes to accommodate this new understanding. Insight is then explained in terms of *self-organized criticality* (SOC) (Gabora, 2001, 2017; Schilling, 2005), a phenomenon wherein, through simple local interactions, complex systems tend to find a critical state poised at the cusp of a transition between order and chaos, from which a single small perturbation occasionally exerts a disproportionately large effect (Bak, Tang, & Weisenfeld, 1988). Thus, while most thoughts have little effect on one's worldview, an idea we call *insightful* is one for which one thought triggers another, which triggers another, and so forth in an avalanche of conceptual change.

Surely, whether one is writing a novel, or composing a symphony, or inventing a new kind of solar panel, this kind of honing process is central to the creative act. Moreover, the ability to hone an idea may have little to do with the ability (or patience) to engage in a futile exercise like coming up with uses for a brick, or things that are red. A refinement on conventional measures of divergent thinking, in which participants indicate what they think are their two most creative answers, and these answers are rated on a 5-point scale, shows good reliability and high predictive validity without the fluency confound (Silvia et al, 2008). However, one could still score highly on this version of the test without having engaged in honing.

Our conception of convergent and divergent thinking may be distorted by our everyday experience in the physical world; because objects in the world exist in different places and have distinct, definite boundaries, it may be difficult to wean ourselves from the intuition that ideas in the mind do as well. It has been argued on the basis of evidence from research on the attributes of associative memory, that the common assumption that creativity involves searching through a space of discrete, separate possibilities, selecting the best, and tweaking it, is misleading (Gabora, 2007, 2010, 2018). This is also what is suggested by research on the formal structure of concepts and their interactions. The goal of the rest of this paper is to, without going into mathematical details, show how this research on concepts points to a new conception of convergent and divergent thinking that resolves the above inconsistencies, and potentially catalyzes a deeper understanding of how the creative process works.

The approach to concepts that I will draw upon is sometimes (somewhat unfortunately) referred to as the *quantum approach* (Aerts, Gabora, & Sozzo, 2013; Aerts & Gabora, 2005; Blutner, Pothos, & Bruza, 2013; Busemeyer & Bruza, 2012; Busemeyer & Wang, 2018;; Gabora, 2001; Gabora & Aerts, 2002; Pothos, Busemeyer, Shiffrin, & Yearsley, 2017). It is called this not because it has anything to do with quantum particles, but because it uses generalizations of mathematical structures originally developed for quantum mechanics. The motivation and rationale for this approach are provided elsewhere (Aerts, Broekaert, Gabora, & Sozzo, 2016b; Bruza, Busemeyer, & Gabora, 2009). For now it is noted that this research by no means aims to reduce cognitive psychology to physics. Rather, much as was the case with other branches of mathematics such as complexity theory and even number theory, structures originally developed by physicists were later found to have applications in other domains. In the quantum approach, concepts are viewed not as fixed representations or identifiers, but as bridges between mind and world that are sensitive to context and that actively participate in the generation of meaning (Gabora, Rosch, & Aerts, 2008).

## Potentiality, Context, and Creative Thought

The gist of the new view of creative thought suggested by concepts research is conveyed by the photograph below of a woodcutting with light shining on it from three different directions, yielding three differently shaped shadows: that of a G, an E, and a B (Figure 1). Though each shadow is different, they are all projections of the same underlying object. We could say that the woodcutting has the *potentiality* to *actualize* different ways, and to actualize in one of these ways requires an *observable* or *context,* in this case, light shining from a particular direction. We can refer to the state of the woodcutting when no light is shining on it as its *ground state*. While it is tempting to assume that a bout of creative thought entails the generation of multiple distinct, separate ideas, there may be a single underlying mental representation that, like the woodcuttings, is ill-defined, and affords some degree of ambiguity in its interpretation. Just because the different sketches of a painting, or prototypes of an invention, take different forms when expressed in the physical world, that doesn't mean they derive from different underlying ideas in the mind. Just as the three shadows of each of the two woodcuttings in Figure 1 are projections of the same underlying object, the sketches or prototypes may be different external realizations of the same underlying idea at different stages of a creative honing process. In other words, these different outputs are different articulations of the idea as it appears looked at from different perspectives. Midway through a creative thought process one may have an inkling of an idea but not yet know whether, or exactly how, it could work. Because it is 'half-baked', it may be more vulnerable to interpretation, meaning that it could appear quite different when looked at from a different perspective.

---

[2] This word is chosen deliberately because it implies that the process is neither entirely top-down nor entirely bottom-up.

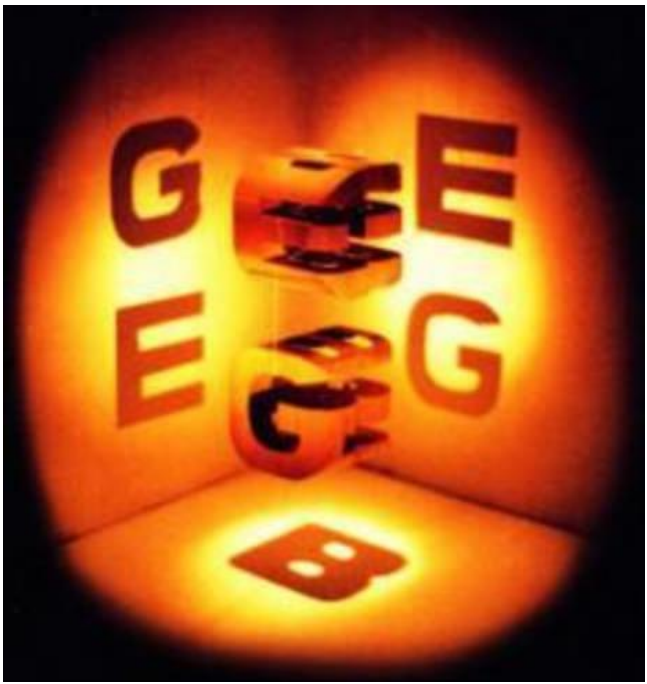

Figure 1: Photograph of ambiguous woodcuttings taken from the front cover of '*Gödel, Escher, Bach: An Eternal Golden Braid*' by Douglas Hofstadter (1979). The top 'triplet' (as he calls them) is not simply a rotated version of the one below it; it is a different shape. (Used with permission.)

Note that the two woodcuttings in Figure 1 have two different shapes, yet they yield the same three shadows. To distinguish the shape of the woodcutting above from the woodcutting below would require that light be shown on them from still more angles, casting shadows that would not look like any particular letters we know. Similarly, the more complex one's unborn creative idea, the more honing steps required to discern its underlying form and whittle it down as needed. Since it has the potential to manifest different ways, we can say that it is a *state of potentiality.*

In the quantum approach to concepts, this kind of potentiality is described as a *superposition state* represented by a vector in a complex Hilbert space. Concepts act as *contexts* for each other that alter how they are experienced; for example, the concept TREE might make you think of a deciduous tree (one with leaves), but in the context CHRISTMAS, you might think of a coniferous tree (one with needles and cones). Each possible context may actualize the potentiality of the concept differently, and these possible actualizations are represented by *basis states*. The actual, existing context is treated as an *observable* that determines how the concept changes in light of this context. It might change in such a way as to alter the weights of certain properties. (For example, 'talks' and 'lives in a cage' are not considered properties of BIRD but they are considered properties of PET BIRD (Hampton, 1987); thus, the context PET is influencing the properties we ascribe to BIRD.) A context can also alter the typicalities of certain exemplars. (As a canonical example, *guppy* is not considered a typical exemplar of PET, nor of FISH, but it is considered a typical exemplar of PET FISH (Osherson & Smith, 1981).) In the absence of any observable—i.e., when a concept is not being viewed from any particular context, or thought about at all—the concept is said to be in a *ground state*. In its ground state there are no properties associated with the concept, but also, there are no properties that are, a priori, excluded from it; thus, you could say it is a state of infinite potentiality. Conceptual change due to the impact of a context is modeled as *collapse* of the vector representing the concept to one of its basis state, as shown in Figure 2.

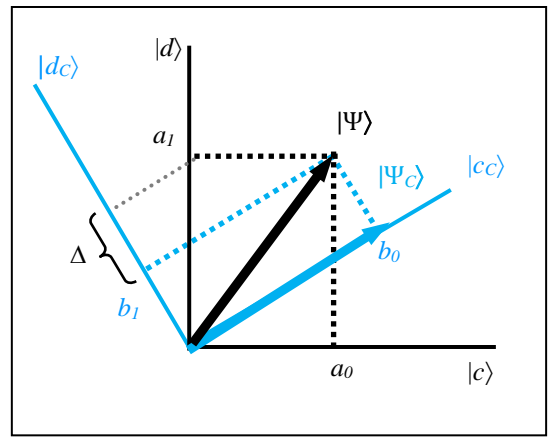

Figure 2: A graphical depiction of a vector $|\Psi\rangle$ representing the concept TREE is shown in black. In the default context, TREE may be more likely to collapse to projection vector $|d\rangle$ which represents DECIDUOUS TREE (tree with leaves) than to collapse to projection vector $|c\rangle$ which represents CONIFEROUS TREE (tree with needles and cones). This can be seen by the fact that subspace $a_0$ is smaller than subspace $a_1$; i.e., $a_0$ is closer to the xy origin than $a_1$. In the context CHRISTMAS, shown in blue, the concept TREE is likely to collapse to the orthogonal projection vector $|c_C\rangle$, representing CONIFEROUS TREE, as shown by the fact that $b_0$ is larger than $b_1$. (After collapse, the projected vector, $|\Psi_C\rangle$, is the same length as the original due to renormalization).

This approach has enabled us to cope with some of the non-compositional ways in which people use concepts—famously said to be the biggest challenge facing cognitive science (Fodor, 1998)—by describing them in terms of effects such as entanglement[3] and interference[4] (Aerts, Sozzo, & Gabora, 2016; Aerts & Sozzo, 2014; Busemeyer & Bruza, 2012). The approach can be applied to concept combinations and more complex compounds of concepts such as decisions (e.g., Busemeyer, Wang, & Townsend, 2006; Yukalov & Sornette 2009) jokes (Gabora & Kitto, 2017), worldviews (Gabora & Aerts, 2009), and creativity (e.g., Gabora & Carbert, 2015). For example, working with data from a study in which participants were asked to rate the typicality of exemplars of a concept for different contexts, and introducing a state-transition threshold, we built a model of how exemplars of a concept arise in divergent versus convergent modes of thought (Veloz, Gabora, Eyjolfson, & Aerts, 2011). By lowering a threshold of allowable deviation from the default context, seemingly atypical exemplars appeared as new possibilities. Honing an idea can be modeled as reiterated collapse, resulting in a change of state of the idea, which induces the conceptual framework to subject the idea to a new context, which in turn brings about a new collapse, and so forth, until the idea is sufficiently robust in the face of new contexts that it no longer undergoes change-of-state (Gabora, 2017). In short, it is becoming possible to move beyond crude conceptions of creative cognition to a more refined understanding that is aligned with and informed by advances in the adjacent area of concepts research.

## Redefining Convergent and Divergent Thought

Let us now see how this can pave the way to a new conception of convergent and divergent thought. There is a relationship between the weights on the properties of a concept in a particular state, and its susceptibility to collapse to any particular new state. For example, if you think about TREE in terms of only its most typical properties such as 'grows in the ground', your next thought may be about something else that grows in the ground, such as a FLOWER. However, if you think about TREE in a way that encompasses not just typical properties such as 'grows in the ground' but also atypical properties, and in particular those implied by the context, your next thought may be about something semantically distant from TREE; for example, a poet might think of a word that rhymes with TREE such as BEE. Recall how, in its ground state, there are no properties associated with a concept, but also, no properties excluded from it. This means that, for any concept there exists *some* context that could come along and make any given property become relevant. The more exotic the context, the more atypical the properties that are evoked, and thus the more unconventional the subsequent thought.

This suggests that in convergent thought an idea is refined by considering compound of concepts in their *conventional contexts*. Because one is not concerned about all the remote ways in which the object of thought could be related to other things, but instead working with it in its most compact form, mental energy is left over for complex operations. This then is why convergent thought is conducive to unearthing relationships of causation, or thinking analytically, as well as simply carrying out rote tasks.

Conversely, in divergent thought one reflects on an idea by considering a particular compound of concepts from *unconventional contexts*. This is conducive to unearthing relationships of correlation, i.e., forging new connections between seemingly unrelated areas, as in analogical thinking. Note that the more unconventional the contexts one calls up, the seemingly less sensible the next thought may be, and therefore the more honing that may be required to coax it into a form that eventually makes sense. It is for this reason that the products of divergent thought (as redefined here to mean thinking of ideas from unconventional contexts) may require extensive honing.

## Implications for Assessment

On the basis of this view of convergent and divergent thinking, let us now re-examine the tests used to assess these constructs. Although the RAT (Mednick, 1968) is used to assess convergent thinking because each question only has one correct answer, to determine the common associate of TANK, TABLE, and HILL you have to think of at least one of these words in a context that is not its default context. For example, unless you are a retailer in the business of selling tank tops you likely interpret the word TANK in terms of its meaning as a military vehicle. Therefore, if we go with the redefinition of convergent thinking as mental operations wherein the contents of thought are viewed from conventional contexts, convergent thought is insufficient to solve the RAT. The RAT is actually more appropriately used as a test of *divergent* thinking. This is consistent with the RAT's wide usage as a test of creativity despite that convergent thought is contrasted with divergent thought and divergent thought is frequently equated with creativity.

---
[3] Entanglement is a phenomenon first encountered in particle physics wherein the state of one entity cannot be described independently of the state of another, and any measurement performed on one influences the other.

[4] Interference is the annihilation of the crest of one wave by the trough of another when they interact.

Since in divergent thinking tasks such as the Alternate Uses task people only reflect upon an idea from unconventional contexts *after* they have generated conventional responses, these tasks only test for divergent thinking during the latter part of the task. Thus it makes sense that, as noted by Beaty and Silvia (2012), this is when the most creative responses occur.

Neither the RAT nor conventional divergent thinking tests assess the capacity to hone an idea in a reiterated manner such that uncertainty decreases to an acceptable level and the idea transitions from ill-defined to well-defined. Amobile's (1982) consensual assessment technique, which involves asking multiple experts to evaluate the creativity of a work, is better in this regard, but it undoubtedly measures not just divergent thinking but what is sometimes called *contextual focus:* the capacity to spontaneously shift between convergent and divergent thought as needed, in response to the situation one is in (Gabora, 2010). What is required is a new approach to creativity testing in which each step in a creative process is broken down into a series of states and contexts, and the type and magnitude of conceptual change from one step to the next are analyzed so as to better understand the interplay of convergent and divergent thinking. Steps in this direction are underway using studies of artmaking (Choi & DiPaola, 2013) and computational models (Bell & Gabora, 2016; DiPaola, 2017; DiPaola, Gabora, & McCaig, 2018; McCaig, DiPaola, & Gabora, 2016), as well as technologies such as functional magnetic resonance imaging (Jung & Vartanian, 2018).

## Conclusions

The constructs of convergent and divergent thought have been around for half a century, and the way they are defined and measured has changed little in that time. Meanwhile, we have made headway in understanding the dynamics of the compounds of concepts that constitute ideas, and in modeling how they interact as one thought gives way to the next. Given the presence of inconsistences in how convergent and divergent thought are conventionally defined and measured, it seems appropriate to revise our understanding of them in light of recent advances in understanding the internal workings of these processes. This paper has shown how formal research on concepts can pave the way to a new way of defining, measuring, and thinking about convergent and divergent thought.

Note that this is not the only potentially fruitful avenue for research yielded by a joining of forces between research on concepts and research on creativity. For example, there are hints that the above-mentioned presence of interference and entanglement effects in empirical studies of conceptual change are related to creativity, but to date this has not been systematically explored. Another direction for future research concerns the role of *incubation:* the idea that setting a creative task aside for a while, or incubating on it, can promote insight. One could model this as letting the idea return to its ground state such that it sheds its coterie of typical properties (and contexts), and taps into its reservoir of infinite potentiality (in the sense that no properties are definitively present nor absent).

Another intriguing prospect this line of inquiry leads to is the following. Creative people are more subject to adoration, as well as social disapproval and even bullying, and it is generally assumed that this is because they violate social norms (Sternberg & Lubart, 1995). However, this may not be the whole story. I have suggested that the creative mind is in the process of honing ambiguous mental forms, and indeed it has long been thought that creative people are particularly comfortable with ambiguity (e.g., Tegano, 1990; but see also, Merrotsy, 2013). This may include ambiguity with respect to how they themselves come across, which in turn may make them more vulnerable to other people's projections. In other words, they may be more subject to misinterpretation, appearing as Gods or Goddesses to some, and as devils to others.

It is hoped that this paper has provided a glimpse of how formal models of concepts can play a key role in the development of a 21$^{st}$ Century understanding of this most human of abilities, the ability to create.

## Acknowledgments

This work was supported by a grant (62R06523) to the author from the Natural Sciences and Engineering Research Council of Canada.